# Equation of State: Manhattan Project Developments and Beyond

Scott Crockett, Franz J. Freibert
Los Alamos National Laboratory
Los Alamos, NM 87545

**Abstract**: The hydrodynamic response of materials under extreme conditions of pressure, temperature and strain is dependent on the equation of state of the matter in all its states of existence. The Trinity plutonium implosion device development required the Los Alamos research community to advance the understanding of equations of state further and faster than ever before. The unpredicted high yield from the Trinity fission device explosion and the push to design the "Super" thermonuclear device initiated 75 years of unprecedented research and technological progress in equation of state development. This article describes the progress made on equation of state development during and since the Manhattan Project at Los Alamos.

## INTRODUCTION TO EQUATION OF STATE AND ITS PURPOSE

The Equation of State (EOS) efforts at Los Alamos have long been responsible for the development and delivery of EOS utilizing the best available experimental and theoretical data. This fact applies broadly to the elements, other metals, alloys and materials in general and actinides in particular due to the Los Alamos National Laboratory mission. As discussed here, the EOS serves as an integral part to solving the hydrodynamic equations used in simulations of high strain rate phenomena, but have significance well beyond that application. A traditional EOS represents the thermodynamic states (pressure $P$, internal energy $E$, and entropy $S$) of matter at a given density $\rho$ (the reciprocal of specific volume $V$) and temperature T. These concepts can be generalized to more complex response of states of matter (i.e., non-isotropic plastic deformation in response to an anisotropic stress tensor, irreversible atomic diffusion, magnetic hysteresis, etc.); however, that requires a deeper technical discussion than will be introduced here. Ensuring a high-quality, first principles informed EOS is essential to predictive hydrodynamic simulations and extreme environment modeling. Small variations in thermodynamic quantities, such as the initial position in ($\rho$, $T$)-space and latent heat or change in enthalpy $\Delta H$ of phase transitions, can have a large propagated effect on simulated trajectories and state endpoints. As will be discussed, the phase and thermophysical properties of matter whether solid, liquid, gas, or plasma and transitions within and interactions amongst these phases, define the EOS for matter and its behavior.

There are numerous technical fields requiring EOS knowledge, in particular, geophysics, planetary sciences, and astrophysics; chemical, aeronautical, and aerospace engineering; and that of nuclear physics, metallurgy, materials engineering, and national security applications which we will examine. When we explore condensed matter behavior in extreme environments, the EOS defines our understanding in a material's physical response to high-strain rate, high temperature, and high pressures conditions. The condensed matter EOS is defined in terms of thermodynamic and physical properties[1], i.e., specific volume, thermal expansion, specific heat, and elastic moduli and sound speeds. Solid state EOS formulations[1], such as those of Grüneisen and Birch-Murnaghan, contain key connections between fundamental thermo-physical properties and response to environmental extremes, whether those extremes are intentionally or accidentally driven.

Developing the fundamental theory behind an EOS requires the interdisciplinary engagement of experts in key topics such as Density Functional Theory (DFT), Quantum Molecular Dynamics, Thermodynamics, and Statistical Mechanics. The development of accurate and predictive EOS theories and advanced modeling leads to reduced uncertainties, improved predictive capabilities from integrated simulations, and accurate physical representations of matter in nature.[2] These theoretical efforts must be closely coupled with experimental efforts in Shock Physics, High Pressure Physics and High Energy Density Physics facilities providing the necessary laboratory data to improve, verify and calibrate theoretic models. A close collaboration between the experimental and theoretical EOS research community has existed since the early days of the Manhattan Project and continues today at Los Alamos National Laboratory and other laboratories around the world.

## BRIEF HISTORY OF EOS AND DEVELOPMENTS LEADING TO THE TRINITY TEST

Prior to World War II, key scientists in the Manhattan Project established many of the fundamental ideas required in understanding EOS theory beyond ideal gases. Many of the models used then and now were established in the late 1800's and early 1900's by Rankine[3], Van der Waals[4], Hugoniot[5,6], Einstein[7], Lindemann[8], Debye[9], Grüneisen[10], Thomas[11] and Fermi[12]. As we examine the history of the EOS effort it will be pointed out how the methods of the past established present modeling philosophies. Early EOS models started from ideal gas approximations which were then extrapolated to include internal degrees of freedom and interactions of the gas molecules[4] to match available experimental data of the time. Limited in temperature, the models would be further extended to include high-temperature thermal electronic responses. These efforts represent foundational principles in thermodynamics, statistical mechanics and quantum mechanics.

Prior to the establishment of Los Alamos, in 1940 Bethe and Teller collaborated on the paper "Deviations from Thermal Equilibrium in Shock Waves"[13] examining oxygen, nitrogen and shock waves in air. Bethe and Teller studied how phenomena such as excitation of molecular vibrations and dissociation connect with statistical equilibrium and how it deviated from known experiments.[13] Continuing in this work, Bethe in 1942 outlined hydrodynamic theory of one-dimensional shock waves in terms of the laws of conservation of mass, momentum and energy.[14] The study was focused on air and water EOS. Bethe elegantly rederived the Rankine-Hugoniot jump conditions.[3,5,6] The thermodynamics of phase transformations and the resultant impact on shock behavior were described by Bethe in detail[14], and this understanding plays an important role for our most modern EOS nearly eight decades later.

When characterizing any new material, certain thermodynamic and physical quantities must be known to properly determine an EOS. In general, one must at a minimum have the following information: a reference density, an atomic weight and number, phase composition, an understanding of the compression response, and phase boundaries (i.e., melt line, solid-solid transformation boundary, etc.). In early 1944, the initial density measurements of gram quantities of the newly produced plutonium wildly varied from 16 g/cc to 20 g/cc.[15] The University of Chicago "Met Lab" and Los Alamos Laboratory metallurgists recognized the fact that multiple allotropes of elemental plutonium existed via density measurements and volumetric dilatometry.[15] The lower density allotrope or δ-phase was malleable, but stabilized at higher temperatures; whereas the higher density allotrope or α-phase was stable at ambient temperatures, but brittle.[16] The pioneering plutonium dilatometry work by Martin, Selmanoff, et al. showed five different allotropes and δ-phase of plutonium surprisingly exhibited a negative thermal expansion[17] (Fig.1). High temperature dilatometry was essential to understanding the temperature at which phase transitions occur, the thermal expansion of each phase, and the density shift between the phases along the one atmosphere isobaric (constant pressure) curve. Also, observable in Fig.1 is thermal expansion hysteresis larger than $\Delta T=150K$ in the γ↔δ phase transformation, which represents the excess strain energy associated with contracting the lattice and increasing the density proceeding from δ-phase (15.92g/cc) to γ-phase (17.14g/cc). Many refinements of these and other thermophysical properties measurements and theory for plutonium have been implemented since then.[18]

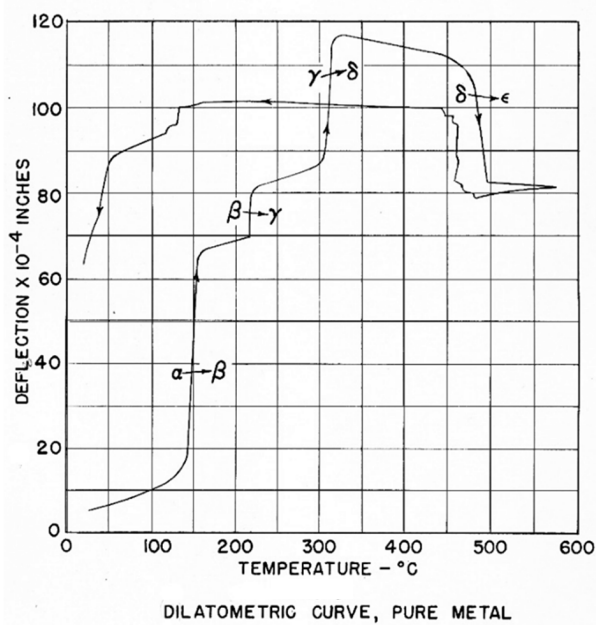

*Figure 1* First dilatometry trace of Martin and Selmanoff[17] for unalloyed plutonium showing five allotropes. Later work on higher purity plutonium revealed the sixth allotrope known as δ'. Reproduced with permission from Los Alamos National Laboratory.

The need for EOS research and hydrodynamic theory and modeling became essential to the Los Alamos mission with the transition from a gun assembled device to a high explosively imploded plutonium device.[19,20] Although the idea of implosion originated in 1942 by Tolman[21], its necessity became evident in early 1944, as it was realized that the hydrodynamics of implosion was necessary to speed the assembly of available plutonium due to its isotopically and elementally impure state and high neutron background to achieve fission yield and avoid the fizzle of predetonation.[22] No one had ever employed high explosives (HE) to assemble a collective of matter before and to do so with precise timing and geometrical uniformity would demand perfect coupling of experimental and theoretical EOS research and development.

The early Manhattan Project implosion plutonium bomb design was based on a hollow plutonium shell for the device core. However, implosion experimentation of 2D hollow objects, such as terminal observation from the HE flash photography of imploding hollow cylinders, indicated an asymmetry of collapse.[23] These asymmetries upon implosion are resultant from hydrodynamic related phenomena such as jetting, spalling, and Rayleigh-Tayler instabilities and led to inconsistent assembly of mass. The Christy Gadget, a tampered solid core device was chosen moving forward based on its design and functional simplicity.[20] As such, concern over accuracies in EOS and hydrodynamic modeling were lessened, but not alleviated.

Early hydrodynamic methods applied to the Christy Gadget development were based on a realistic multiphase EOS model that assumed imploding material changes discontinuously; however, complications arose when a multiphase EOS was employed. The numerical solution of the partial differential equation describing the hydrodynamics of implosion were too difficult for Los Alamos computing staff to hand calculate. Considerable amount of effort was put into the multiphase model, but the results proved very difficult to interpret. In the spring of 1944, newly purchased IBM machines were configured to solve the implosion problem, a theory effort led by Teller in Theory (T-1) Group.[23] Teller and collaborators derived approximations to result in a simplified EOS used for IBM calculation and first results of the implosion simulations were extremely satisfactory. Implementation of multiphase EOS models under extreme conditions[24] was delayed until modern computational capabilities and experimental validation and verification could support such developments.

Teller and Ulam submitted "An Equation of State in the Condensed Phase for Arbitrary Pressures and Moderate Temperatures" in September 1944.[25] In this paper, they approximated the EOS by splitting internal energy into two parts: a low temperature thermal ion (vibrational) part which is temperature dependent and a compressional energy part which is volume dependent via Birch's parameterization. The basic approach allowed them to approximate the low-pressure shock response of the material and had assumptions similar to the Mie-Gruneisen theory. By November 1944, preliminary interpolated curves representing EOS for Al, Cd, Fe, and U were produced[26] and by January 1945 Metropolis generated zero temperature compression "cold" curves for a number of metals leveraging equations of Fermi and Thomas.[27] This Fermi and Thomas model, now considered the simplest example of a Density Functional Theory, was published shortly after Schrodinger's paper introducing his wave equation.[28] Having the effective cold curve enables an approximate lower bound of the compression energy for a given state of matter density.

A memo dated April 16, 1945 detailed the first of Bridgman's compression experiments on plutonium and was followed up with additional measurements reported April 25, 1945 of different samples up to ~0.1Mbar.[29,30] Other materials data were reported soon afterward, e.g., Peierls discussed shock wave experiments on Al, Cd, Cu, Fe, U in May 1945.[31] The original work was performed using a smooth-bore gun, due to geometric constraints and shockwave interactions exhibited by the HE lens systems[19] under development by Goranson, et al.[32] By the July 1945 Trinity Test and subsequent Nagasaki blast, T-Division devoted significant time to the hydrodynamics interpretation of the blast measurements and to the radiation hydrodynamics of the implosion fission bomb.[33] During the same period, shock and material velocities had been measured in six metals and six non-metals,[34] and with this work, Ancho Canyon in Los Alamos became the birth place of modern shock physics. Shock dynamics and the effects on materials were studied from the beginning of the Manhattan Project as discussed by Marshak[35] and summarized by Taylor[36]. An extensive discussion on the history of the experimental shock program at Los Alamos was written by Taylor.[36] According to Taylor, gas-guns used for measurements of Hugoniots only came into use by early 1955 and later in 1958. Hints of plutonium experiments (restricted to the Nevada Test Site) occurred around 1955.

In September 1945, Keller, et al.[37] made refinements to the EOS from the work of Metropolis[27] which aimed to better incorporate temperature effects. These temperature effects leveraged the cold curve response from the Thomas-Fermi-Dirac (TFD)[38] and Bethe-Marshak approach for finite temperature to the electron density. The ion model follows the previous discussed work of Teller and Ulam.[25] Special care was made in interpolating between the Bridgman data which goes to 0.1 Mbar and the TFD calculation which started at ~75Mbar. They used a simple analytic form where the results compared favorably with the initial work of Christy.[26] The first comparisons were made of uranium shock data[32] to 1/3 Mbar showing only a 7% deviation from the experiment. Present day work differs mainly in that the experimental data and theoretical calculations overlap more closely, thereby reducing or eliminating intermediate regions of interpolation. It is now standard practice for condensed matter EOS determination to initiate a three-term decomposition of Helmholtz free energy $F$ into

$$F = E_S + F_V + E_{e^-} \quad (1)$$

for which $E_S$ is the $T=0K$ static lattice cold curve typically derived from DFT, $F_V$ is the vibrational energy of the ions from thermal excitation as elucidated by the Debye model, and $E_{e^-}$ is the energy of free electron density from thermal excitation as described in TFD theory. [1,2]

## POST-TRINITY EOS RESEARCH

Soon after the Trinity Test, Los Alamos was reorganized to address the next greatest driver for the laboratory which was the development of the "Super", a thermonuclear fusion device. The July 16, 1945 Trinity Test led to the interest concerning radiation hydrodynamics and the implosion fission bomb.[33] The higher than anticipated yield of the Trinity explosion[33,39] resulted in a revitalization of earlier speculations regarding the simplifying assumptions and disregard for radiation built into the original implosion efficiency calculations. Researchers began to take notice of these "simplifications" as interesting areas of research. For example, it has long been recognized that the Rankine-Hugoniot relations for shock waves and the empirical relation between the shock-wave and particle velocities define an incomplete thermodynamic description of the states along the Hugoniot curve.[40] This thermodynamic limitation drove shock experiments diagnostic development to measure temperature under shock conditions in metals which led to such challenging experiments as the use of thermocouples for rapid temperature determination.[36] On Aug 26, 1947, Mayer suggested the use of a short duration burst of x-rays to observe the shock and detonation front in materials[41] from which to estimate temperature excursion. The application of flash x-ray radiography techniques has been utilized to observe shockwave phenomena for six decades; however, the ability to make accurate estimate of temperature using the flash x-ray technique are only possible in recent years at facilities such as the Los Alamos National Laboratory DARHT facility. Theory developments led to the 1956 Cowan extension of temperature range in Thomas-Fermi-Dirac models[42] which furthered this model to allow for calculations of temperature into the keV range. The methods which Cowan used were simplified by Liberman around the late 1960s and this adaptation of the TFD model is what we currently use for free electron density contributions to EOS.[43]

Other EOS developments post-Trinity have included focus on metal, nuclear materials and minerals. By 1953, Cowan reanalyzed previous work[37] and made improvement to the aluminum and uranium EOS by accounting for error in the 24ST aluminum used in the initial shock experiments. As the understanding of the EOS of standard materials improves, the validation of new experimental platforms improves and uncertainty in measurements is reduced. Continued refinement and improved fidelity in our simulation capabilities requires higher precision EOS and hence improved experimental resolution. Fickett and Cowan also explored the idea of analyzing shock compression of sintered versus cast samples[44] to better understand the impact of varied microstructures and morphologies, which was a means to study porous materials. This and related work has great value in geophysics[1] and seismic based nuclear non-proliferation monitoring and treaty verification. Further utilization of EOS during this time included the Bethe and Tait 1956 fast reactor safety analysis which considered the internal energy of the system determined by the fission rate in the molten liquid fuel and the EOS of the fuel material, specifying the relationship between the nuclear aspects of excursion and the dynamic response of the core.[45]

The first publicly available post-war review of shock physics and EOS was published in 1958 by Rice, et al. entitled "Compression of Solids by Strong Shock Waves".[46] This work provided a review of research and technological advancement that had occurred to date within the shock physics community. In particular, Rice, et al. discussed the evolution of the experimental methods used to measure and produce strong shock waves and provided the Hugoniot curves for twenty-seven materials. The experiments utilized a flash gap approach with a photographic method to measure the shock fronts. The authors also compare the shock data with the static work of Bridgman. Rice, et al. summarize the capabilities that existed between the experimental and theory efforts.

Post-war shock and EOS research represented the most-rapid expansion of knowledge in this field. That precipitous growth and accumulation of data brought recognition of the need to consolidate and document those data sets for use in computing. In the 1960s, Cowan established *Maple*, the first tabulated EOS database; and in early 1970s, Barnes was instrumental in creating the *SESAME I* database. The *SESAME* database was intended to be a standardization of equation of state, opacity, and conductivity information. Barnes and others leveraged the now ample amount of shock data and recalibrated the *Maple* tables. Barnes et al. subtracted the zero temperature cold curves and replaced them with calibrated cold curves that matched experiment. The thermal components (ion and electron) remained the same. By this time, many of the capabilities to generate the thermal components of an EOS had been lost, at which point the EOS project was rebuilt by Barnes, with Kerley as the project lead. For much of the 70's, 80's, and 90's, Abdallah, Albers, Bennett, Boettger, Dowell, Holian, Johnson, Liberman, Lyons, Rood, Straub, Barnes and Kerley contributed to the creation of the *SESAME II* database.[47,48] *SESAME II* was made available to the world in 1979. The researchers listed here utilized the models developed by our project founders and solidified the methodology we employ today.[49]

### Modern EOS Methods

A modern EOS model is built by leveraging data collected over experimental conditions ranging from ambient,

static compression, and shock regimes, and by developing an integrated theoretical approach to ensure dataset inclusion. These modern models cover compressions of 0 to $10^6$ volumetric strains and temperatures from 0 to $10^9$ Kelvins. Once optimized across multiple data sets, the model forms naturally extend to known thermodynamic limits. We start, however, with ambient data along the one-atmosphere isobar. The EOS relevant thermophysical properties data includes information for the reference density, thermal expansion, specific heat, and bulk moduli. X-ray diffraction measures the initial crystal structure and density. Dilatometry measures the thermal expansion. Resonant Ultrasound Spectroscopy is used to measure the adiabatic bulk modulus. Calorimetry is a measurement of the enthalpy and specific heat. Experimental methods also provide the temperatures of phase transitions (solid-solid, solid-liquid, liquid-gas, solid-gas). These isobaric data provide constraints to the thermal components of an EOS model. Then theoretical calculations are used for constraining our models in regions where data is often absent. For that we rely on DFT, Quantum Molecular Dynamics, and Quantum Monte Carlo (QMC) calculations. These methods are computationally intensive, but better match experimental results for most materials describable by a multiphase EOS.[24] Such a modern multiphase EOS generated at Los Alamos National Laboratory for aluminium is shown in Fig.2. Other modern EOS models include that of the Lawrence Livermore National Laboratory PURGATORIO, a novel implementation of the INFERNO equation of state physical model.[50]

Examining the compression of materials at zero temperature, DFT calculations are leveraged with Diamond Anvil Cell (DAC) measurements and then extended to high compression by matching TFD calculations. DAC is a measurement of static compression typically measured at room temperature. By heating the DAC, it can be leveraged to measure phase boundaries to moderate temperature. Quasi-isentropic compression experiments, pioneered by Sandia National Laboratories on the Z-Machine in the early 2000s, provides a revolutionary means by which to measure compression of high Z (atomic number) materials including actinides to extreme pressures. Isentropic compression, which leaves the entropy of the system constant, achieves higher compression and lower temperature states than single shocks. These experiments can be combined with shock (Hugoniot) experiments allowing access to unique regions of thermodynamic space as shown in Fig.2.

Shock compression experiments test the EOS of a material through the Rankine-Hugoniot relations, based on conservation of mass, momentum, and energy. These relations are temperature independent, and as a result even a poor-quality EOS can produce the correct Hugoniot as this is not a unique relationship. The difference in quality of EOS lies within how the EOS predicts the derivative quantities off the Hugoniots such as the sound velocity and release response, hence the importance of combining shock data with other sources in order to calibrate an EOS. Shocks can be obtained by multiple platforms. Gas-guns, laser, high-explosives, magnetic, and nuclear are all means of delivering energy into a sample to produce high strain rate compressive states.

EOS modelers must be careful when interpreting experimental data of solids. So-called EOS measurements in a solid, whether by dynamic or static means, often contain deviatoric effects such as strength, kinetics, or other non-hydrostatic mechanical or non-equilibrium responses that are accounted for not in the EOS, but by other physics models. Therefore, comparisons with experiment must leave room for other models to come in and provide a complete picture.

## CONCLUSIONS

The Manhattan Project efforts conducted at Los Alamos represent the most prolific era in the history of equation of state development. The shift from a gun assembled plutonium device to an implosion assembled plutonium device demanded a rapid advancement in equation of state experiment and theory to support the hydrodynamic simulations necessary for device development. However, over the intervening 75 years, many of the original EOS simplifications employed to yield a tractable Trinity device implosion simulation have been set aside as significant improvements and innovations have been developed. These developments range from computable multiphase equations of state, to insitu density diagnostics, to an array of experimental platforms on which to conduct EOS studies.

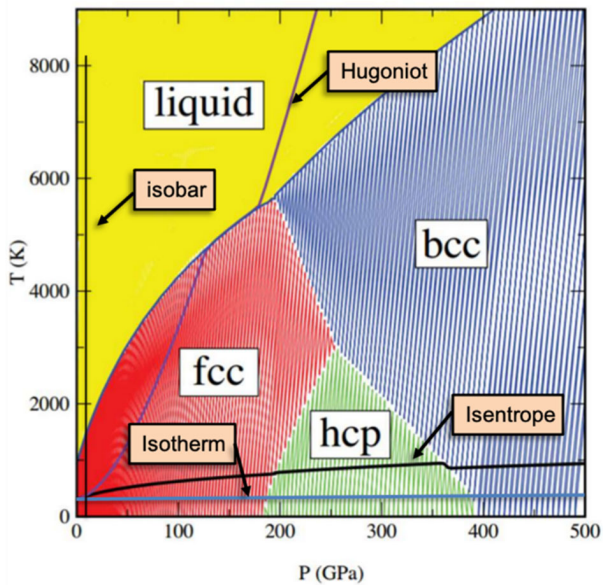

*Figure 2* The phase diagram for aluminum overlaid with key thermal dynamics curves. The isobar (constant pressure), isotherm (constant temperature), adiabat or isentrope (constant entropy), and the Hugoniot are standard thermal dynamics paths that can be measured via experiments. The labels fcc, hcp and bcc represent regions of face-centered cubic, hexagonal close-packed and body-centered cubic phase stability, respectively.

**ACKNOWLEDGEMENTS**

We want to thank J.D. Johnson, R. Hixson, C. Greeff, E. Chisolm, D. Wallace and J. Martz for the discussions and assistance with this article. This work was supported by the U.S. Department of Energy through the Los Alamos National Laboratory. Los Alamos National Laboratory is operated by Triad National Security, LLC, for the National Nuclear Security Administration of U.S. Department of Energy under Contract No. 89233218CNA000001.